# Vanishing quantum confinement enables bright and thermally excited multi-carrier emission from semiconductor nanocrystals

Tjom Arens[1,2], Sander J.W. Vonk[3], A. Willem Vlasblom[1], Margarita Samoli[4], Daniel Vanmaekelbergh[2], Pieter Geiregat[4,5], Zeger Hens[4,5] and Freddy T. Rabouw[1,2]

[1]*Soft Condensed Matter & Biophysics, Debye Institute for Nanomaterials Science, Utrecht University, Princetonplein 1, 3584CC Utrecht, The Netherlands*
[2]*Condensed Matter & Interfaces, Debye Institute for Nanomaterials Science, Utrecht University, Princetonplein 1, 3584CC Utrecht, The Netherlands*
[3]*Optical Materials Engineering Laboratory, Department of Mechanical and Process Engineering, ETH Zurich, 8092 Zurich, Switzerland*
[4]*Physics and Chemistry of Nanostructures, Department of Chemistry, Gent University, Krijgslaan 281, 9000 Gent, Belgium*
[5]*NOLIMITS, Core Facility for Non-Linear Microscopy and Spectroscopy, Gent University, Krijgslaan 281, 9000 Gent, Belgium*

**Abstract**

Recently, nanocrystals in the regime of vanishing quantum confinement—termed bulk nanocrystals (BNCs)—have demonstrated remarkable optical gain characteristics. While their high-power lasing performance was demonstrated convincingly, the photophysics at low and intermediate powers—where charge-carrier populations are discrete—remain unexplored. Using single-photon avalanche diode (SPAD) array technology, we resolve the dynamics and energetics of six multi-carrier excited states in individual CdSe/CdS core/shell BNCs, containing up to four electrons and two holes. Each state exhibits bimodal emission, indicative of thermal equilibrium between closely spaced electron and hole levels, confirmed via temperature-dependent single-particle measurements. Quantification of radiative and nonradiative decay channels reveals strongly suppressed Auger recombination through both the negative- and positive-trion pathways. We present a model that combines statistical scaling of rate constants with Fermi–Dirac thermal occupations of electron and hole levels, bridging the transitional regime between quantum-confined and bulk nanocrystals, and providing a comprehensive framework for understanding this emerging class of materials.

**Introduction**

Semiconductor nanocrystals (NCs) are used as light-emitting components for commercial displays and hold great promise for lighting and laser applications.[1] Over the past three decades, design strategies such as core–shell architectures[2,3] and alloyed interfaces[4–6] have been optimised to enhance photoluminescence efficiencies by mitigating surface-related losses and suppressing nonradiative Auger recombination. Auger processes, which quench multi-carrier excited states under strong excitation, have long posed a key limitation for achieving efficient and sustained lasing in NCs, restricting early demonstrations to short-pulse excitation and high thresholds.[7] These challenges were partially overcome by selectively delocalising carriers into the shell[8,9] or by grading the confinement potential at the core–shell interface[4–6].

More recently, attention has shifted from shell engineering to the effect of core size. Large NCs with diameters comparable to or exceeding the Bohr radius have demonstrated efficient spontaneous and stimulated emission from multi-carrier states, indicative of strongly suppressed Auger recombination.[10–13] Depending on composition, such particles are termed quantum shells[14,15] if the innermost core is a high-bandgap material, or bulk nanocrystals[11,12] (BNCs) if the core is the material with the lowest bandgap. These materials show long gain lifetimes, consistent with their ability to sustain optical gain under quasi-continuous-wave excitation.[11,12,16] The lasing behaviour and the gain spectra of BNCs were described using a bulk semiconductor model with continuous charge-carrier densities and a continuous density of energy levels.[17–19] Yet, the structured gain spectra hinted at discrete energy levels.[11] Moreover, charge-carrier densities are discrete around or below the gain threshold, which is the regime most relevant to lighting or laser operation. The behavior of NCs with vanishing quantum confinement, where the energy levels are dense but discrete and where charge-carrier densities are low but discrete, remains unexplored.

Addressing this challenge requires methods capable of resolving discrete multi-carrier states in individual NCs. Ensemble transient-absorption studies have provided valuable insights[20], but quantitative interpretation is complicated by particle-to-particle variations in energetics and dynamics, as well as spontaneous charging[21] and the simultaneous generation of a variety of multi-carrier states. Single-particle spectroscopy can overcome these limitations by directly resolving multi-carrier states[19,22–24], including charged configurations that arise spontaneously[25] or can be introduced intentionally[26,27] in laser design strategies.

In this work, we isolate and characterise the energetics and dynamics of six multi-carrier states in Cd chalcogenide BNCs with an average core diameter of 10.5 nm, placing them in the transitional regime between bulk and quantum-confined. We resolve multi-carrier states in individual BNCs with a number of holes up to 2 and electrons up to 4, using a method termed cascade spectroscopy[19,28] or heralded spectroscopy[23,24] on a single-photon avalanche diode (SPAD) array detector. The photoluminescence (PL) spectra are bimodal, with a relative intensity scaling with the number of charge carriers, highlighting thermal occupation of higher electron and hole levels. The expected temperature dependence of level occupations is confirmed with temperature-dependent single-particle measurements using specialised microscope substrates capable of localised heating[29]. We quantify radiative and nonradiative decay pathways of each multi-carrier state. Nonradiative Auger losses are spectacularly reduced compared to smaller NCs, as highlighted by the emission efficiency as high as 60% for a highly excited state with 2 holes and 4 electrons. Finally, we develop a unifying model of discrete electron and hole levels with Fermi–Dirac occupation statistics to describe the size regime between quantum confinement and bulk.

**Resolving the emission spectra of different excited states in a single nanocrystal**

We prepare our sample by spin-coating a strongly diluted solution of CdSe/CdS core/shell BNCs (core diameter: 10.1 ± 1.2 nm; total diameter: 15.4 ± 1.5 nm; mean ± standard deviation as determined from electron microscopy; Supplementary Fig. 1; cf. Bohr diameter: 6.8 nm[30]) onto a glass coverslip. The coverslip is fixed onto a microscope slide under a nitrogen atmosphere using an airtight adhesive spacer. The exclusion of oxygen stabilises excess electrons in the conduction band and facilitates negatively charged multi-carrier states.[31] We perform pulsed-excitation single-BNC experiments with a 405-nm diode laser operating at a repetition rate of 1 MHz and a fluence corresponding to an average number of excitations per pulse of $\mu \sim 0.2$ (Supplementary Note 2.1). The emission from a single BNC is spectrally dispersed by a transmission grating onto a one-dimensional SPAD array, which together act as a spectrometer setup with single-photon time-tagging capabilities and a spectral resolution of 1.84 nm (Fig. 1a and Supplementary Fig. 2).[23,24] We verify the presence of only one active emitter in the laser focus using a time-gating method (Supplementary Note 3).

The emission intensity and excited-state lifetime fluctuate, caused by spontaneous charging and discharging of the BNC (Supplementary Fig. 4).[2,32] We isolate the moments when the BNC is charged by analysing the fluorescence-lifetime–intensity distribution (FLID; Supplementary Note 4.1), a 2D histogram that shows the number of photons in 40 ms time bins and the corresponding average fluorescence lifetime, as shown in Fig. 1b. We distinguish three emissive states, which we ascribe to different degrees of negative charging of the BNC promoted by the oxygen-free atmosphere.[33,34] We identify the highest-intensity state (300 cts / 40 ms; 50 ns; red) as the neutral BNC and the two states with progressively lower intensity (265 and 240 cts / 40 ms) and shorter lifetime (26 and 18 ns) as the singly (orange) and doubly (yellow) negatively charged states. The observation of bright charged states indicates that Auger quenching in these BNCs is weak. This behaviour is consistent with the large particle size, which reduces Auger rates owing to volume scaling[35,36], and with the long gain lifetimes that point to suppressed Auger recombination[11,12].

Our time-tagged spectral data allow the PL spectra of different multi-carrier states to be reconstructed. Each photon recorded during the experiment is assigned to a different charge state based on the FLID analysis described above (Supplementary Note 4.1). Under low-excitation conditions, these photons originate mostly from singly (as opposed to doubly) excited states. Hence, the integrated spectra of the three charge states (Fig. 1d) are those of the neutral exciton (1 electron + 1 hole), singly charged exciton (2 electrons + 1 hole), and doubly charged exciton (3 electrons + 1 hole). A fraction of the photons originates from doubly excited states, which we isolate from cascade events in our photon stream—more specifically, laser pulses followed by two photon-detection events. As the recorded signal originates from a single BNC, the first and second photon of a cascade must be emitted in succession, i.e. by decay from the doubly excited state via the singly excited state to the ground state (Supplementary Note 4.2).[19,23] Fig. 1e shows the PL spectra of the doubly excited states with different negative charges. We will refer to these as the neutral biexciton (2 electrons + 2 holes), singly charged biexciton (3 electrons + 2 holes), and doubly charged biexciton (4 electrons + 2 holes).

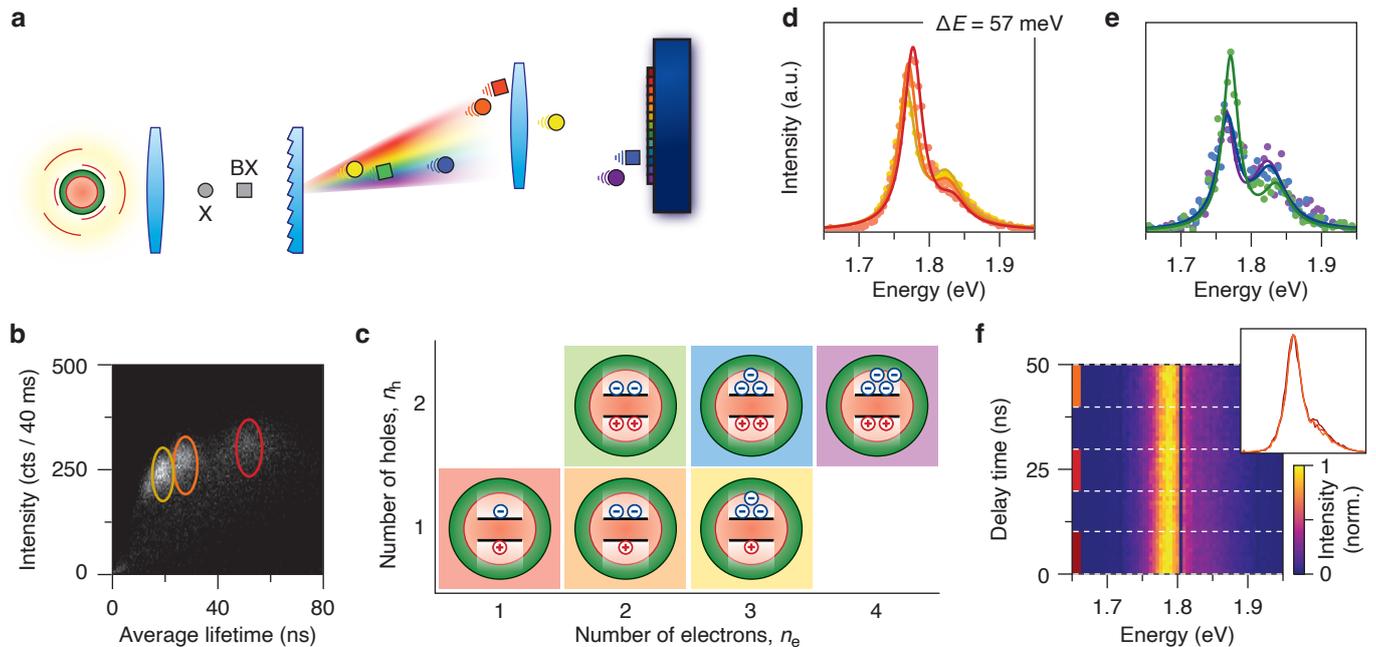

**Fig. 1 | Experimental setup and emission dynamics of a single bulk nanocrystal. a,** Schematic of the single bulk nanocrystal (BNC) experiment. A single BNC is excited by a pulsed laser; emitted photons are spectrally dispersed and time-tagged by a linear single-photon avalanche diode (SPAD) array. By post-selecting the first photon of each photon pair following a single excitation pulse, biexciton emission is isolated. **b,** Fluorescence-lifetime intensity distribution (FLID), showing fluctuations in average lifetime and intensity over a total experiment, caused by spontaneous charging and discharging of the BNC. Three distinct emissive states are identified based on their average lifetime: neutral (red), singly negatively charged (orange), and doubly negatively charged (yellow). **c,** Combining the post-selection of biexciton photons with charge-state identification enables the isolation of six distinct excited states, corresponding to configurations with one or two holes and up to four electrons. **d,** Reconstructed photoluminescence (PL) spectra of the neutral exciton (red), singly charged exciton (= trion; orange), and doubly charged exciton (yellow). **e,** Same as d, but for the neutral biexciton (green), singly charged biexciton (blue), and doubly charged biexciton (purple). The rise in the relative intensity of the higher-energy peak with the number of excess electrons is attributed to increased population of the electron P level. **f,** Normalised time-resolved emission map showing spectral evolution of the neutral exciton emission as a function of delay time. The inset displays the integrated spectra over delay ranges of 0–10, 20–30 and 40–50 ns. At early times (0–10 ns), the higher-energy P-level contribution is enhanced, consistent with the presence of the neutral biexciton, which has a shorter lifetime than the exciton and contributes primarily at short delays. At later times, when only exciton emission remains, the spectral shape stabilizes. These results indicate that the two emitting levels are thermally coupled.

The spectra of Figs. 1d,e are bimodal for all multi-carrier states, with a dominant low-energy transition at 1.78 eV and a weaker high-energy transition at 1.83 eV. The low-energy peak redshifts by 7 meV and 10 meV with respect to the neutral state for the singly and doubly charged exciton states, respectively. Notably, the intensity ratio between the two peaks changes: the relative intensity of the high-energy transition increases with the number of excess electrons for both the exciton and the biexciton emissions. These spectra are qualitatively different from room-temperature spectra of smaller quantum dots (QDs). Single-QD emission spectra sometimes show a tail on the red side of the main peak, but these are well understood as phonon replicas.[37,38] In contrast, the side peak observed here is on the blue side.[11,12]

Asymmetric emission spectra have been observed in ensemble-scale studies of BNCs[10,11] and quantum shells[14]. The bimodal structure is more distinct in our single-BNC spectra because of reduced spectral broadening. Lv et al.[10] ascribed the blue emission shoulder to excited particle-in-a-box electron and hole levels for their cubic BNCs. The equivalent in our spherical BNCs would be the P-symmetry levels. Indeed, the experimental energy splitting of $\Delta E = 57$ meV is close to the expected energy difference between $1S_{3/2}1S_e$ and the $1P_{3/2}1P_e$ states of 82 meV from a simple single-carrier effective-mass model (Supplementary equation (S8)). To check that the $1P_{3/2}1P_e$ emission is due to thermal excitation of charge carriers—rather than transient emission as charge carriers thermalise—we construct a time-resolved emission map (Fig. 1f for the neutral exciton; Supplementary Fig. 6 for all exciton states with enlarged insets). The constant spectral shape over delay times up to 50 ns is consistent with thermal coupling between excited states strongly outpacing recombination from either state.

The scaling of the intensity ratio between the two emission peaks with increasing number of charge carriers (Figs. 1d,e) is as expected from the fermionic nature of charge carriers in BNCs. The degeneracy of the lowest 1S levels is low: $g_e = 2$ and $g_h = 4$ for the electron and hole, respectively. As the number of charge carriers increases, the thermal occupations of the 1S and 1P do not grow at the same pace. Instead, the fermion occupation of a level is limited by the degeneracy, as described by Fermi–Dirac statistics. Quantitative analysis of Fermi–Dirac occupations will follow later.

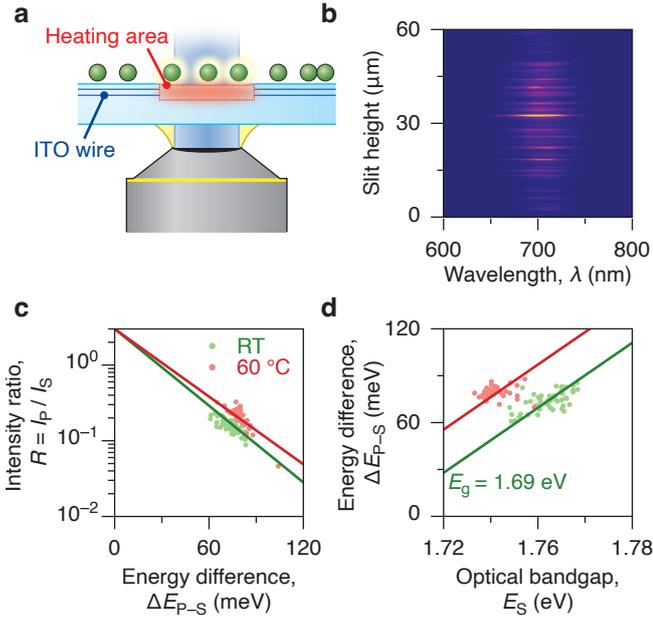

**Fig. 2 | Temperature-dependent multi-particle spectroscopy. a,** Schematic of the experimental setup used for widefield, high-throughput temperature-controlled spectroscopy of single BNCs. Dozens of BNCs are excited simultaneously using widefield illumination through an oil-immersion objective, while an Interherence Vaheat substrate heats them. Emission is spectrally dispersed using a slit and grating and imaged onto an electron-multiplying CCD camera. **b,** Time-averaged PL spectra of multiple single BNCs, which are the streaks at different heights on the camera image. Each BNC spectrum is fitted to extract the SS transition energy $E_S$, the energy separation between the SS and PP transition $\Delta E$ and the P-to-S fluorescence intensity ratio $R = I_P/I_S$. **c,** Scatter plot of the experimental $R$ versus $\Delta E$ at room temperature (20 °C; green) and 60 °C (red). Each data point is an individual BNC. The intensity ratio increases for smaller $\Delta E$ and higher temperatures, consistent with thermally activated P emission. Solid lines are the expected intensity ratio as a function of $\Delta E$ according to a Boltzmann model with a degeneracy ratio of $g_P/g_S = 3$. **d,** Same as c but for the experimental $\Delta E$ versus $E_S$. Solid lines are fits with a slope fixed by a particle-in-a-box-model and the bulk bandgap as an optimisable parameter.

To confirm thermal occupation of P-symmetry energy levels we raise the temperature during single-particle spectroscopy (Supplementary Note 2.2). This is made possible with Interherence Vaheat coverslips[29] with localised heating up to 100 °C through resistive heating of transparent conductive wiring (Fig. 2a). We excite the sample in widefield and guide the emission from dozens of individual BNCs via a slit and a reflective grating onto a 2D electron-multiplying CCD camera (Fig. 2b). To ensure characterisation of the neutral exciton state (1 electron + 1 hole), the experiments were conducted in air and under low-light continuous-wave excitation. We obtain values for the energy of the SS transition $E_S$, the energy difference between the SS and PP transition $\Delta E$ and the P-to-S intensity ratio $R = I_P/I_S$, both at room temperature (20˚C) and at 60 °C by fitting Gaussians to the spectrum of each BNC (Supplementary Note 5).

Variations in the intensity ratio between BNCs correlate with the energy splitting $\Delta E$ and increase at 60 °C compared to room temperature (Fig. 2c). This is exactly as expected for thermal coupling between S and P levels. Unintentional variations between nominally identical BNCs in terms of particle size thus present a convenient test of thermal occupation. For the exciton state—with only a single carrier per band—the fermionic nature of charge carriers does not affect level occupations and Boltzmann statistics are expected. Indeed, the data points in Fig. 2c match the Boltzmann model

$$R = \frac{g_P}{g_S} \exp\left(-\frac{\Delta E_{P-S}}{k_B T}\right) \tag{1}$$

where $g_P/g_S = 3$ is the degeneracy ratio between bright SS and PP exciton states (Supplementary Note 6.2), $k_B$ the Boltzmann constant and $T$ is the temperature (293 K or 333 K). The good match of equation (1) to the data without any optimisable parameters indicate equal oscillator strength of SS and PP transitions.

We also observe a correlation between $\Delta E_{P-S}$ and $E_S$ (Fig. 2d). This is expected for energy variations due to size differences, as the particle-in-a-spherical-box model predicts a $V^{-2/3}$ scaling for both $\Delta E_{P-S}$ and $E_S$ (Supplementary Note 6.3). Specifically, a relation of

$$\Delta E_{P-S} = \frac{\chi_{11}^2 - \pi^2}{\pi^2}(E_S - E_g) \tag{2}$$

is expected, where $\chi_{11} = 4.49$ is the first zero of the spherical Bessel function of order 1 and $E_g$ is the bulk bandgap energy of CdSe. We fit equation (2) to the data in Fig. 2d, fixing the slope of the line and optimising only $E_g$. The fitted room-temperature value of $E_g = 1.69$ eV is slightly lower than literature values (1.74–1.76 eV),[11,39–41] which can be understood from the 60-meV Stokes shift for this batch of BNCs (Supplementary Fig. 7). The fitted bandgap value at 60 °C is 1.67 eV. A decrease of the bandgap energy on the order of 10 meV between room temperature and 60 °C is consistent with expansion of the semiconductor lattice and increased electron–phonon interactions, as empirically described by the Varshni equation.[42]

## Characterising the intermediate regime between bulk and quantum confinement

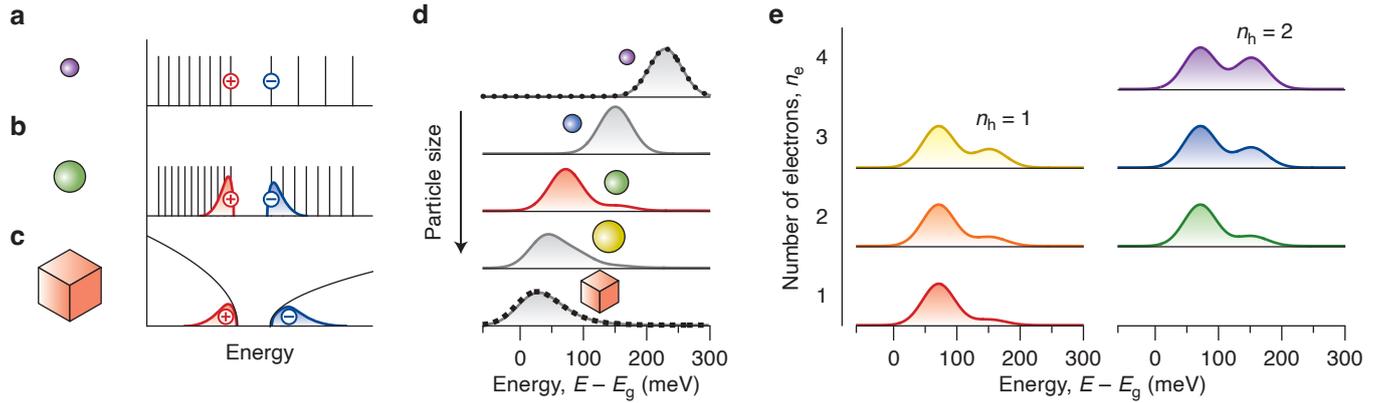

**Fig. 3 | Energetics of multi-carrier excited states in CdSe bulk nanocrystals. a–c,** Schematic representation of the energy-level separation and state filling across three particle size regimes: **(a)** strongly quantum-confined, **(b)** intermediate, and **(c)** bulk. In the intermediate regime, discrete energy levels remain, but their populations follow Fermi–Dirac statistics. **d,** Model calculations for the emission of the neutral exciton state, based on the effective-mass model for energy level spacing and Boltzmann distribution for level occupation. Different graphs represent different NC sizes ranging from 3× smaller (purple) to 10× larger (red) than our BNCs. The dashed lines on the top and bottom spectra are the predictions of conventional models for strongly confined NCs and for a bulk semiconductor, respectively. Each calculated spectrum is broadened by 25 meV. Our model connects the two regimes of bulk and quantum-confined NCs. **e,** Extension of the model to include multiple electrons and/or holes per NC, by incorporating the fermionic nature of the charge carriers. In agreement with experimental observations, the calculated spectra show an increase in the relative intensity of high-energy peaks with increasing electron number.

Our observations lead to a conceptual understanding of the energetics of BNCs, which naturally links the description of charge carriers in strongly confined NCs and in a bulk semiconductor (Figs. 3a–c). Strongly confined NCs have discrete energy levels and charge carriers occupy only the lowest level in the exciton ground state (Fig. 3a). Higher energy levels are occupied only at high charge carrier densities or transiently following non-resonant excitation. On the other extreme of the size range, bulk materials have a continuous density of states and their occupation is described by Fermi–Dirac statistics (Fig. 3c). The transitional regime—that of BNCs—is well described with a discrete-level model including thermal excitation of S electrons and holes to P levels (Fig. 3b).

Fig. 3d presents theoretical emission spectra of the neutral exciton state for increasing particle size, where each spectrum is broadened by 25 meV (Supplementary Note 6.4). These simple calculations neglect Coulomb interactions between carriers and assume equal oscillator strength for the $1S_{3/2}1S_e$ and $1P_{3/2}1P_e$ excitons. The calculated spectrum for the smallest particles (3× smaller than our BNCs; purple) matches the prediction of a conventional model for strongly confined NCs (dashed black line). The calculated spectrum for the largest particles (10× larger than our BNCs; red) matches the prediction of a conventional bulk model (dashed black line). Our model reproduces the expected behavior in both limiting regimes and captures the emergence of discrete spectral features in the intermediate, transitional size range. Thermal excitation was thus the missing link for the description of the transitional size regime.

Describing thermal excitations for multi-carrier states beyond the neutral exciton requires Fermi–Dirac statistics, accounting for the fermionic nature of electrons and holes. This leads to increased occupation of higher excited states at elevated charge-carrier densities (Supplementary Note 7). Consistent with experimental observations (Fig. 1d,e), the calculated relative intensity of the high-energy peaks increases with the number of electrons in the NC (Figs. 3e).

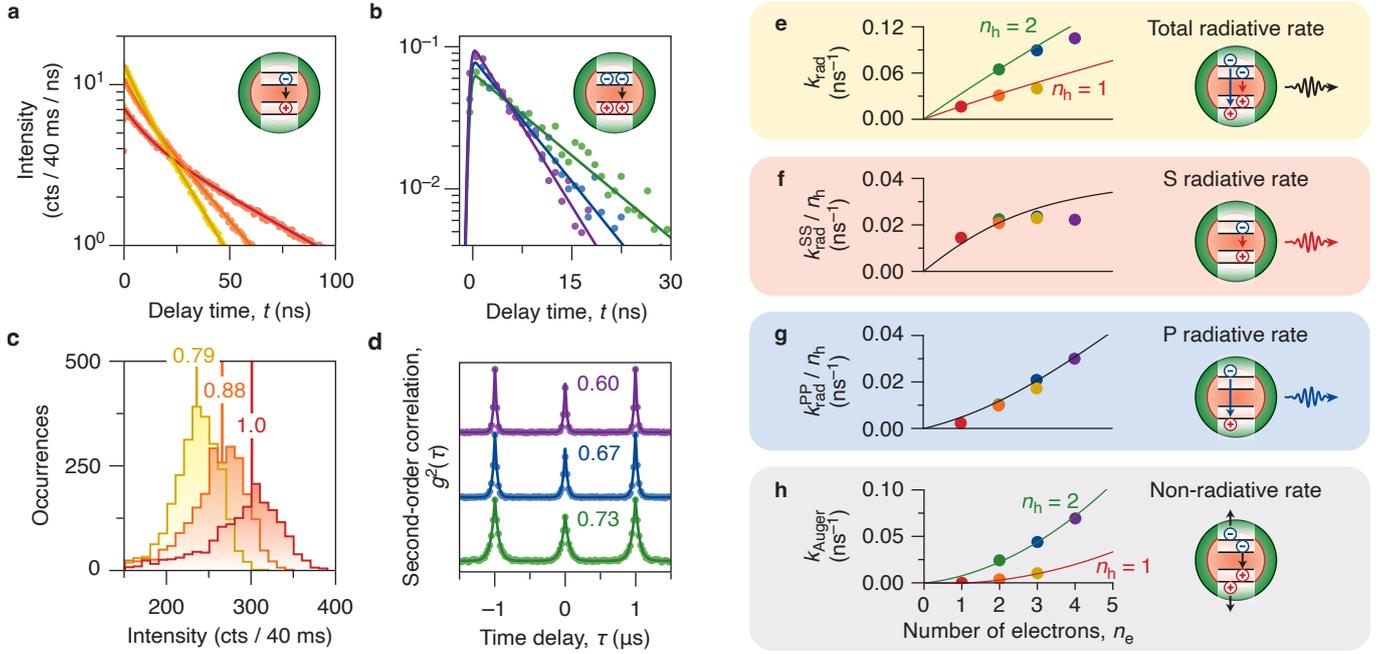

**Fig. 4 | Statistical scaling validated over many excited states in a single BNC. a,** Photoluminescence decay curves of the neutral (red), singly charged (orange), and doubly charged (yellow) states. (Charged) exciton lifetimes are extracted from the slow component of a biexponential fit, while the fast component is attributed to the (charged) biexciton. **b,** Photoluminescence decay curves of the neutral (green), singly charged (blue), and doubly charged (purple) biexciton states, reconstructed by analysing photon cascades. **c,** Histogram of photon count rates from the neutral (red), singly charged (orange), and doubly charged (yellow) states, the solid lines show the centre of a Gaussian fit. Assuming a unity quantum efficiency of the neutral exciton, we estimate the efficiencies of the other states from the relative count rates.[43] **d,** Second-order correlation functions of the neutral (green), singly charged (blue), and doubly charged (purple) state of the BNC. We estimate the (un)charged biexciton quantum efficiencies from the peak height in the correlation function and the (un)charged exciton quantum efficiencies determined in panel c. **(e)** The total radiative decay rate $k_{rad}$ as a function of the number of electrons $n_e$ in a multi-carrier state. Solid lines: expected trends following Eq. 6 for states with $n_h = 1$ (red) and $n_h = 2$ (green), taking $B'$ such that the theoretical and experimental $k_{rad}$ match for the biexciton. **(f)** The SS radiative transition rate (Eq. 7) scaled with $n_h$ and as a function of $n_e$. Solid: expected trend according to Eq. 8, using $B'$ from panel e. **(g)** Same as f, but for the PP radiative transition rate. **h,** The total non-radiative decay rate as a function of $n_e$. The solid lines are the expected Auger decay rates for the various charge configurations (Eq. 9) calculated using Auger coefficients $C^- = 0.0017$ and $C^+ = 0.0040$ ns$^{-1}$ for the negative and positive trion pathways, respectively.

Thermal excitation is also expected to be reflected in the recombination dynamics. We quantify the recombination rates of the excited states identified in Fig. 1. Specifically, we reconstruct the (cascaded) time-averaged decay curves of the exciton and biexciton for each charged state (Fig. 4a,b). We find lifetimes (Supplementary Note 4.3) for the neutral, singly charged and doubly charged biexciton of $\tau_{BX} = 11.2$ ns, $\tau_{BX-} = 7.5$ ns, and $\tau_{BX2-} = 5.8$ ns, respectively. We then extract the exciton lifetimes from a biexponential fit (Supplementary Note 4.4) to the total decay curves, fixing the short lifetime component at the biexciton lifetime. The lifetimes for the neutral, singly charged and doubly charged exciton are $\tau_X = 59$ ns, $\tau_{X-} = 28$ ns, and $\tau_{X2-} = 20$ ns, respectively.

To distinguish radiative and non-radiative decay contributions, we first determine the quantum efficiencies of the excited states from their relative count rates and intensity-correlation analysis (Supplementary Note 4.5). The quantum efficiency of the neutral exciton is assumed to be 100%, because in single-BNC measurements we observe only the bright fraction of BNCs of the synthesis batch.[43] The quantum efficiencies of the charged exciton states are obtained from the relative count rates compared to the neutral exciton (Fig. 4c), giving $\eta_{X-} = 88\%$ for the singly charged exciton and $\eta_{X2-} = 79\%$ for the doubly charged exciton. Under low excitation fluences, the biexciton-to-exciton quantum efficiency ratio can be determined from the intensity-correlation function $g^{(2)}$ (Fig. 4d):[44]

$$\frac{\eta_{BX}}{\eta_X} = \frac{g^{(2)}(0)}{g^{(2)}(\pm T)} \tag{3}$$

where $g^{(2)}(0)$ is the amplitude of the zero-delay peak and $g^{(2)}(\pm T)$ is the average amplitude of the side peaks. Accounting for the finite quantum efficiencies of charged excitons (Fig. 4c), we determine the quantum efficiencies of the neutral, singly charged and doubly charged biexciton at $\eta_{BX} = 73\%$, $\eta_{BX-} = 67\%$ and $\eta_{BX2-} = 60\%$, respectively. These high quantum efficiency values of multi-carrier states are consistent with suppressed Auger recombination in large NCs, but are exceptionally high here exceeding 50% even for states with up to 6 charge carriers.[15,33,45] We then convert the lifetimes and the quantum efficiencies into radiative $k_{i,r}$ and non-radiative $k_{i,nr}$ decay rates for each state $i$:

$$k_{i,\text{rad}} = \frac{\eta_i}{\tau_i} \quad \text{and} \quad k_{i,\text{nonrad}} = \frac{1-\eta_i}{\tau_i}. \tag{4}$$

The radiative and non-radiative decay rates of six multi-carrier states are an excellent data set to test the validity of statistical scaling on recombination rates and the possible influence of thermal occupation. In bulk, where energy levels are effectively degenerate and thermal effects are negligible, the radiative decay rate scales simply with the number of electrons $n_e$ and holes $n_h$:

$$k_{rad} = B n_e n_h. \tag{5}$$

In BNCs, where energy levels remain discrete, the scaling is determined by the expected number of carriers of opposite charge thermally occupying states of the same symmetry:

$$k_{rad} = B' n_e^S n_h^S + \frac{1}{3} B' n_e^P n_h^P \tag{6}$$

where $n_i^j$ is the occupation of charge carrier $i$ in level $j$ and $B'$ is chosen such that the calculated total radiative decay rate of the uncharged biexciton matches the experimental value. The exact solution for discrete carrier numbers can be obtained via our microstate model (Supplementary Note 7.1). Instead, to produce continuous lines (Fig. 4e), we use the well-known Fermi–Dirac distribution function $n_i^j = 1/[e^{(E_i - \mu)/k_B T} + 1]$, which is a decent approximation even for systems with down to 1–4 fermions (Supplementary Note 7.2 and reference [46]). We isolate the radiative decay rates through the SS and PP transitions:

$$k_{rad}^{SS} = k_{rad} f^{SS} \quad \text{and} \quad k_{rad}^{PP} = k_{rad} f^{PP} \tag{7}$$

where $f^{SS}$ and $f^{PP}$ are the fraction of emitted photons in the low-energy and high-energy peak, respectively. These fractions are obtained from a least-squares fit of the PL spectrum using a double Lorentzian model (Figs. 1d,e). The experimental $k_{rad}^{SS}$ and $k_{rad}^{PP}$ (Figs. 4f,g) follow the expected trends

$$k_{rad}^{SS} = B' n_e^S n_h^S \quad \text{and} \quad k_{rad}^{PP} = \frac{1}{3} B' n_e^P n_h^P \tag{8}$$

where $n_i^j$ is the Fermi–Dirac occupation of charge carrier $i$ in level $j$. The prefactors $B'$ are equal for the two model lines in Figs. 4f,g, again demonstrating that the SS and PP transitions have similar oscillator strength (Supplementary Note 7.3). A deviation between experiment and model is apparent for $k_{rad}^{SS}$ at $n_e = 3$ and 4, which we ascribe to reduced electron–hole overlap due to electron–electron repulsion.

The scaling of the non-radiative decay rates (Fig. 4h) demonstrates that they are due to three-carrier recombination process, most likely Auger recombination. We fit the non-radiative decay rates to

$$k_{Auger} = C^- n_e n_h (n_e - 1) + C^+ n_e n_h (n_h - 1) \tag{9}$$

where $C^-$ and $C^+$ are prefactors for the negative- and positive-trion Auger pathway, respectively.[34] The good match of this expression with the experiment indicates that S and P levels are active in Auger processes at a similar rate. Simpler models used for bulk semiconductors, which disregard the discrete number of charge carriers per BNC, fail to match the experimental data (Supplementary Fig. 9), highlighting the effects of discrete and finite carrier numbers on non-radiative Auger decay. The parameter values are $B' = 58.3$ μs$^{-1}$, $C^- = 1.68$ μs$^{-1}$, and $C^+ = 3.96$ μs$^{-1}$ if the charge densities are expressed in terms of number of charge carriers; or $B' = 1.13 \times 10^{-10}$ cm$^3$ s$^{-1}$, $C^- = 6.40 \times 10^{-30}$ cm$^6$ s$^{-1}$, and $C^+ = 1.50 \times 10^{-29}$ cm$^6$ s$^{-1}$ if converted to units of number of charge carriers per unit of volume, considering the entire BNC core/shell volume.

**Discussion**

Our conceptual framework for the energetics of BNCs naturally incorporates elements from existing models: the discrete states from strongly confined NCs with the Fermi–Dirac thermal occupations of a macroscopic semiconductor. Other materials with weak exciton binding likely exhibit similar thermal excitations and scaling of transition rates with carrier density in the transitional size regime, including CdS[12], InP, InAs, and PbS. In contrast, the photophysics of materials such as lead–halide perovskites is determined by bound excitons, even for larger NCs.[47] A distribution of oscillator strength over an extended spectral range may be an undesired aspect of BNCs—compared to smaller NCs—for application as a laser gain material or a phosphor material. On the other hand, the finite degeneracy of the lowest-energy state, similar to smaller NCs, makes it easier to achieve population inversion compared to bulk materials. New applications of thermally excited emission could include optical cooling[48] or background-free anti-Stokes microscopy[49].

The emission efficiencies of multi-carrier states in BNCs are spectacularly high, as quantified in our work for isolated states with up to four electrons. Qualitatively, such high emission efficiencies are in line with the volume dependence of undesired Auger recombination.[33,35,45] Despite the large BNC volumes, statistical-scaling relations still reflect the involvement of discrete charge carriers rather than a continuous charge carrier density. Fig. 5 summarizes the emission efficiencies, lifetimes and P emission fractions of the six multi-carrier states studied experimentally (Figs. 5a,c,e) and the values predicted by our comprehensive model and extrapolated to even higher multi-carrier states [Figs. 5b,d,f; equations (8) and (9)]. The excellent agreement highlights the strength of our model, as it predicts the efficiencies of numerous excited states. The high emission efficiencies and long lifetimes of multi-carrier states make BNCs an ideal solution-processable materials class for optoelectronic applications.[11,12] Auger quenching is limited even for states with as many as four electrons, highlighting the potential for zero-threshold lasing by intentional charging.[26,27] Our framework provides a predictive and intuitive tool for interpreting the optical behaviour of large NCs, as they become increasingly relevant for modern optoelectronic and photonic applications.

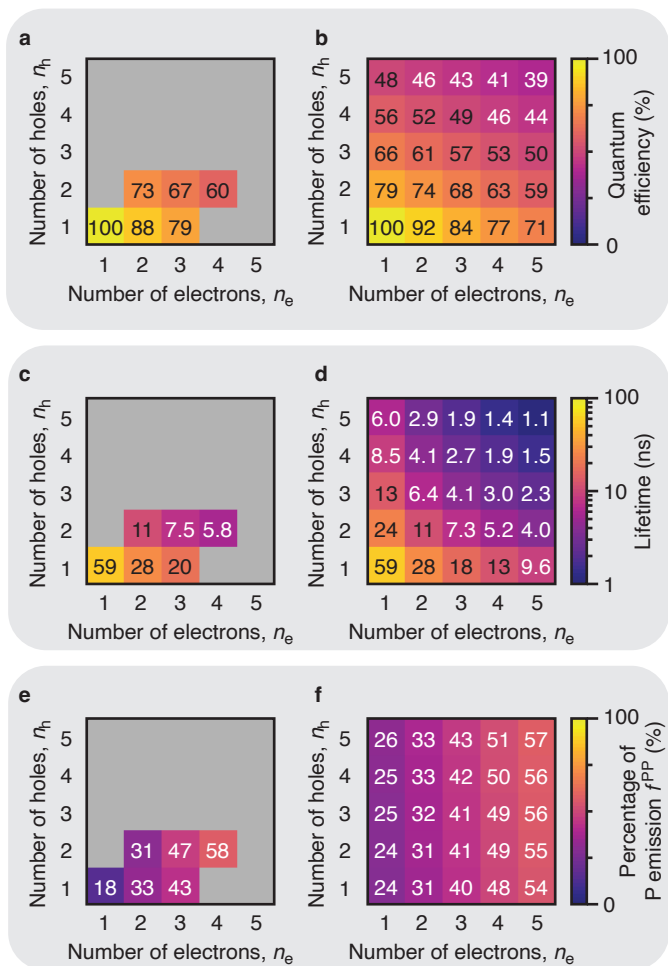

**Fig. 5 | Emission efficiencies and lifetimes of multi-carrier states in bulk nanocrystals. a,** Experimental quantum efficiencies of multi-carrier states in the BNC studied in Figs. 1,4, as a function of the number of electrons $n_e$ and the number of holes $n_h$. **b,** Same, but calculated using the discrete microstate model (Supplementary Notes 7.1, 7.3 and 7.4) and using the radiative and Auger recombination rates determined in Fig. 4. The model is extrapolated to multi-carrier states with up to 5 electrons and 5 holes. **c,d,** and **e,f,** Same as **a,b**, but showing the lifetime and the P emission fraction of multi-carrier states, respectively.

Supplementary Information for:

# Vanishing quantum confinement enables bright and thermally excited multi-carrier emission from semiconductor nanocrystals

Tjom Arens[1,2], Sander J.W. Vonk[3], A. Willem Vlasblom[1], Margarita Samoli[4], Daniel Vanmaekelbergh[2], Pieter Geiregat[4,5], Zeger Hens[4,5] and Freddy T. Rabouw[1,2]

[1]*Soft Condensed Matter & Biophysics, Debye Institute for Nanomaterials Science, Utrecht University, Princetonplein 1, 3584CC Utrecht, The Netherlands*

[2]*Condensed Matter & Interfaces, Debye Institute for Nanomaterials Science, Utrecht University, Princetonplein 1, 3584CC Utrecht, The Netherlands*

[3]*Optical Materials Engineering Laboratory, Department of Mechanical and Process Engineering, ETH Zurich, 8092 Zurich, Switzerland*

[4]*Physics and Chemistry of Nanostructures, Department of Chemistry, Gent University, Krijgslaan 281, 9000 Gent, Belgium*

[5]*NOLIMITS, Core Facility for Non-Linear Microscopy and Spectroscopy, Gent University, Krijgslaan 281, 9000 Gent, Belgium*

## Supplementary Note 1: Synthesis and characterisation of bulk nanocrystals

**Chemicals**

Cadmium oxide (CdO, 99.5+%), toluene ($C_7H_8$, 99.9%), methanol ($CH_3OH$, 99.8+%) and 2-propanol ($C_3H_8O$, 99.8%) were purchased from Chemlab Analytical. Elemental selenium (Se, 200 mesh) and cadmium acetate dihydrate ($Cd(CH_3COO)_2 \cdot H_2O$, 98% for analysis) were purchased from Acros Organics. Elemental sulfur (S, ≥ 99.5%) was purchased from Sigma Aldrich and tri-n-octylphosphine ($C_{24}H_{51}P$, min. 97%) was purchased from Strem Chemicals. Squalane ($C_{30}H_{62}$, 98%) and oleic acid ($C_{18}H_{34}O_2$, tech. 90%) were purchased from Thermo Fischer Scientific. All chemicals were used as is with no further purifications.

**Slow-injection synthesis of CdSe/CdS BNCs**

The 10.5/15.5 nm particles were synthesized using the same procedure described in reference [1].

The Cd-oleate (0.50 M) solution was prepared as follows: in a 50 mL three-neck flask, 1.024 g cadmium oxide (8 mmol), 8 mL oleic acid (25 mmol) and 8 mL squalane were added and degassed for an hour at 110 °C. After degassing, the flask was set under inert conditions and heated to 300 °C for around 5-10 min. As soon as the solution turned clear, it was let to naturally cool down to <120 °C and then collected.

The TOP-Se (0.50 M) and TOP-S (0.50 M) solutions were prepared in separate 20 mL vials inside a $N_2$-filled glovebox. 10 mL tri-n-octylphosphine (22.4 mmol) were added to 0.395 g selenium (5 mmol) and 0.160 g sulphur (5 mmol), respectively. The vials were left to stir at 90 °C for around 30 min until the solutions turned clear.

For the slow - injection synthesis, 8 mL squalane were added in a 50-ml three-neck flask and degassed at 110 °C for an hour. After degassing, the flask was filled with $N_2$ and heated to 340 °C. In the meantime, an equimolar mixture of pre-heated Cd-oleate and TOP-Se was prepared in a glovebox. The mixture was inserted in a 5-ml syringe and slowly injected into the flask using a 2 mL/hour rate. The total injection time is modified as needed to produce the desired core size. For the 10.5 nm CdSe core, a total of 0.5 mL Cd-oleate and 0.5 mL TOP-Se were mixed and injected over the course of 60 min. After the injection, the flask was held at 340 °C for around 10 min to allow for full precursor consumption.

To form the CdS shell with a 2.5 nm thickness, a 5-ml syringe filled with an equimolar solution of 0.75 mL Cd-oleate and 0.75 mL TOP-S was similarly prepared. In a one-pot process, the second solution was injected at 340 °C with the same rate into the reaction flask over the course of 90 min. After the injection, the flask was held for an additional 30 min at 340 °C to allow for full precursor consumption.

For the purification, the cooled-down crude medium was split into four centrifugation tubes containing a 2:1 ratio mix of 2-propanol and methanol (around 15 mL in each tube). The tubes were then centrifuged at 5000 rpm for 10 min to precipitate the particles. This process was repeated for a total of two cycles. The particles were then re-dispersed in 3 mL toluene for further use.

An acetate ligand post-treatment was finally implemented to help increase the particle luminescence. For this, 100 μL of oleic acid and 0.0268 g of grinded cadmium acetate dihydrate were added in a 20 mL vial along with the full purified BNC batch. The partially open vial was then sonicated at 80 °C for 90 min. After sonication, the vial was centrifuged at 5000 rpm for 15 min to remove any unreacted salts and coagulated particles. The sur natant was collected and filtered in a new vial, where 6 mL 2-propanol were added to start a mild purification step. The particles were crashed by centrifuging at 5000 rpm for 3 min and finally re-dispersed in 3 mL toluene.

**Characterisation**

The core/shell dimensions of the BNCs were then verified through bright-field TEM imaging. The images were obtained on a JEOL JEM-1011 microscope equipped with a thermionic gun operating at 100 kV accelerating voltage. Image processing was done using ImageJ software. Particle sizing was performed using a Python script based on a binary thresholding model that distinguishes particles from the background by pixel intensity. After image filtering and thresholding, the particles were identified, size-filtered, and scaled to nm. Assuming spherical geometry, the average diameters were calculated, and size distributions were generated from over 300 particles.

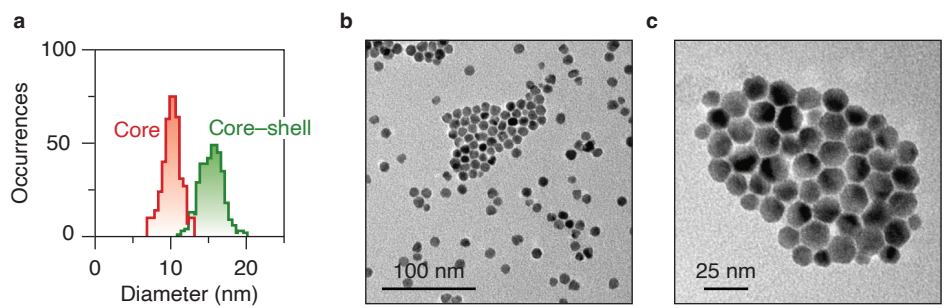

**Supplementary Fig. 1 | Particle size analysis of CdSe/CdS BNCs. a,** Size distributions of CdSe cores (red) and CdSe/CdS core–shell BNCs (green) obtained from TEM. Average diameters: 10.1 ± 1.2 nm (core) and 15.4 ± 1.5 nm (core–shell). **b,c,** Representative TEM images of CdSe/CdS BNCs.

**Supplementary Note 2:** Experimental setup

**Supplementary Note 2.1:** Cascade spectroscopy

Single bulk nanocrystals (BNC) measurements were conducted using a custom-built microscope setup based on a Nikon-Ti-U inverted microscope. The sample was prepared by spin-coating a highly diluted solution (1:20,000 dilution compared to the stock) onto a glass coverslip. This coverslip was then mounted onto a microscope slide under a nitrogen atmosphere using an airtight adhesive spacer, effectively excluding oxygen and stabilising excess electrons in the BNC.

BNC excitation was achieved using a pulsed 405-nm laser (Picoquant D-C 405, controlled by Picoquant PDL 800-D laser driver operating at 1 MHz), which was guided to the sample by a dichroic mirror (425 nm, Thorlabs DMLP425R). For single-particle measurements, the light was focused through a high-numerical-aperture oil-immersion objective (Nikon CFI Plan Apochromat Lambda 100×, NA 1.45). Emission from the sample was collected through the same objective and spectrally dispersed using a transmission grating (Thorlabs, 300 lines/mm). With an achromatic aspherical lens (Edmund, $f = 50$ mm) the Fourier plane of the transmission grating was then imaged onto a one-dimensional single-photon avalanche photodiode (SPAD) array consisting of 320 pixels (SPAD Lambda, Pi Imaging), achieving a spectral resolution of 1.84 nm/pixel. To eliminate residual laser light, a long-pass filter (430 nm, Bright line) was placed in the detection path. Based on the count rate relative to a calibrated setup[2] with two single-pixel SPADs, we determine a detection efficiency of approximately 4%. The excitation rate $\mu$ for the experiments in the main text is ~0.2 excitations per laser pulse.

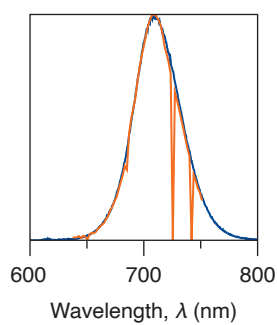

**Supplementary Fig. 2 | Wavelength calibration of the time-correlated spectrometer.** To calibrate the wavelength recorded on each pixel of our time-correlated spectrometer, we dropcasted a 10× diluted BNC stock solution on a glass coverslip and measured the emission of a thick layer on a glass substrate. A flipper mirror was used to alternate the emission path between an electron-multiplying CCD (Andor iXon Ultra 888) with a calibrated spectrograph (Andor Kymera 193i) and the SPAD array. By overlaying the photoluminescence (PL) spectra from both detectors, we assigned wavelengths to individual SPAD pixels. Two hot pixels with extreme dark-count rates were identified and rejected from all further analysis in the single-BNC experiments. This procedure yielded a spectral resolution of 1.84 nm per pixel for the SPAD setup.

**Supplementary Note 2.2:** Temperature-dependent multi-particle spectroscopy

For the temperature-dependent multi-particle experiments, the excitation path was modified by inserting an additional 20-cm-focal-length lens to focus the laser onto the back focal plane of the objective, enabling widefield illumination. The sample preparation followed the same protocol as for the single-particle measurements, with the exception that the BNCs were spin-coated onto a Vaheat temperature-controlled substrate (Interherence, maximum temperature 100 °C) instead of a glass coverslip.

To minimise multiexciton generation and suppress charging effects, the measurements were performed under low-intensity continuous-wave (CW) excitation at 405 nm and under ambient conditions. Prior to data acquisition at higher temperatures, the sample was heated (to 60 °C) and allowed to thermally stabilise for 5 minutes.

The emitted photoluminescence was dispersed by a spectrograph and recorded using an electron-multiplying CCD camera. Time-averaged PL spectra were acquired using the following settings: EM gain 300, slit width 50 µm, 100 accumulations, and 0.1 s exposure time per frame.

## Supplementary Note 3: Blind-time analysis

Conventionally, the observation of anti-bunching, defined as $g^{(2)}(0) < 0.5$ in the normalised correlation function $g^{(2)}(\tau)$, is regarded proof of a single emitter in the detection volume of a spectroscopy setup. However, the high multiexciton emission efficiencies of BNCs cause reduced anti-bunching, such that $g^{(2)}(0) > 0.5$ even for a single BNC. To assess whether the emission originates from a single BNC or a cluster of emitters, we therefore employed a blind-time analysis based on $g^{(2)}$. This method uses the difference in the characteristic lifetimes of the two photons in a cascade event, a biexciton photon followed by an exciton photon. Since the first photon of a cascade (BX) has a relatively short lifetime compared to a photon emitted from an exciton state (X), applying a short blind period ($\Delta$) after a laser pulse rejects coincidence counts originating from a cascade more strongly than coincidences from two independently emitted exciton photons.[3]

If we study a single BNC with high biexciton emission efficiency, the coincidence counts in the zero-delay peak are due to cascade emissions. Because the exciton photon always follows the biexciton photon the number of photon pairs in the zero-delay peak is proportional to the number of biexcitons that are emitted outside the blind period:

$$g^{(2)}(0, \Delta) \propto \exp\left(-\frac{\Delta}{\tau_{BX}}\right) \tag{S1}$$

In contrast, the side peaks (at ±1 pulse repetition period) reflect uncorrelated exciton photons from subsequent excitation cycles—in the low excitation regime. The number of photon pairs in the side peak as a function of the blind window scales as:

$$g^{(2)}(\pm T, \Delta) \propto \left[\exp\left(-\frac{\Delta}{\tau_X}\right)\right]^2 = \exp\left(-\frac{2\Delta}{\tau_X}\right) \tag{S2}$$

By comparing the decay behaviour of correlated photon pairs as a function of $\Delta$, one can infer whether the observed emission is consistent with a single BNC or a collection of emitters (Supplementary Fig. 3).

Fitting the peak amplitude as a function of blind time for the zero-delay and side peak yields a biexciton lifetime $\tau_{BX}$ = 10.8 ns and an exciton lifetime $\tau_X$ = 47.2 ns, respectively. These values are similar to the values that we obtained from the time-averaged decay curves ($\tau_{BX}$ = 11.2 ns and $\tau_X$ = 58.9 ns), indicating that the collected emission originated from a single BNC.

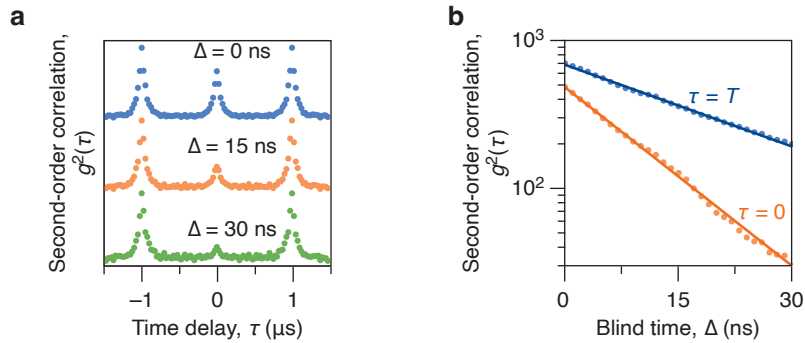

**Supplementary Fig. 3 | Determining the number of emitters via blind-time analysis. a,** Second-order correlation function $g^{(2)}(\tau)$ of the neutral exciton, calculated while omitting all photons arriving within $\Delta$ = 0 ns (blue), 15 ns (orange), and 30 ns (green) after each laser pulse. Increasing the blind time rejects photon pairs due to fast biexciton emission from the biexciton cascade more strongly compared to pairs from consecutive exciton photons. As a result, the zero-delay peak is expected to decrease more rapidly than the side peaks. **b,** Amplitude of the zero-delay peak (orange) and the side peak (blue) as a function of blind time. The zero-delay peak decays more rapidly than the side peak, consistent with the shorter lifetime of the biexciton. This behaviour indicates that the emission originates from a single emitter rather than a cluster of emitters.

**Supplementary Note 4:** State selection and excited-state characterisation

To isolate six multi-carrier states from the continuous photon stream emitted by a single BNC, we identify the number of excess charges in the BNC for each 40-ms time bin of the experiment by lifetime thresholding and subsequently distinguish between exciton and biexciton photons through selection of cascaded emission events.

**Supplementary Note 4.1:** Charged-state selection

To obtain the intensity and lifetime for each 40-ms time bin over the course of our experiment, we applied maximum-likelihood estimation (MLE) to fit the corresponding decay curves, accounting for the Poissonian statistics of photon counting. This procedure yielded both the average fluorescence lifetime and emission intensity per bin, which were subsequently used to construct the fluorescence-lifetime–intensity distribution (FLID; Fig. 1b in the main text).

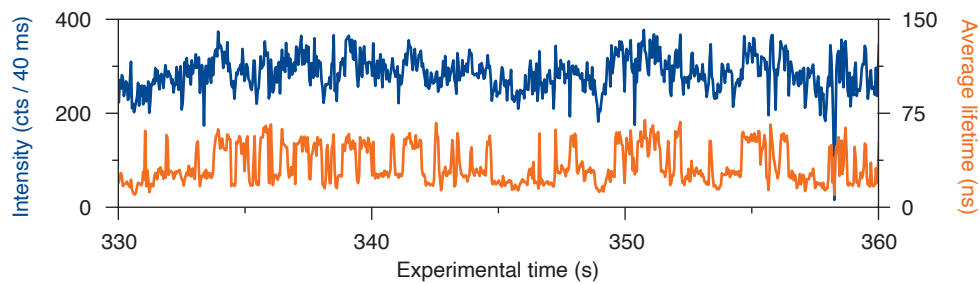

**Supplementary Fig. 4 | Blinking trace showing fluctuations in intensity and lifetime due to spontaneous charging.** A segment of the experiment from the main text is shown. The intensity trace exhibits multiple emission levels without distinct boundaries, whereas the corresponding average lifetime trace reveals clear transitions between different charge states.

Periods of BNC charging are characterised by a lower count rate and shorter excited-state lifetime, because of increased non-radiative and radiative recombination pathways (Supplementary Fig. 4). Previous studies on smaller QDs often made the distinction easy because the efficiency of the charged (OFF) state was relatively low compared to the neutral (ON) state. Our BNCs, however, have charged-state emission efficiencies above 70%. This makes separation of charged and neutral periods based on intensity almost impossible. We therefore separate the states based on their average lifetime, instead. For clean selection of signals from neutral, singly charged, and doubly charged states and to reduce the effect of spectral diffusion, we use strict lifetime thresholding. Specifically, in the histogram of the average lifetimes (Supplementary Fig. 5), the neutral state is selected using a lifetime range of 45–57 ns the singly charged state with 23–29 ns, and doubly charged state with 15–21 ns. In our experiments the laser fluence remains constant and we assume the absorption cross section to remain constant. We can therefore determine the emission efficiency of a charged state from the average counts (per 40 ms) compared to the neutral state, which we assume to have unity emission efficiency.[4]

With each photon assigned to specific charged states, we can reconstruct the photoluminescence (PL) spectra and time-averaged PL decay curves. We normalise each decay curve to the time spent (number of 40-ms time bins) in a certain state and each spectrum to its emission maximum.

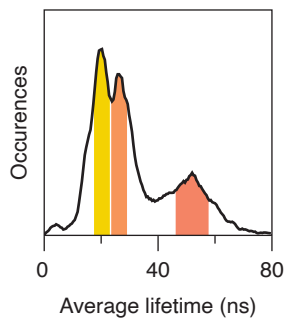

**Supplementary Fig. 5 | Lifetime histogram revealing multiple emitting states.** Histogram of the average lifetimes of photons within 40-ms time bins. Four distinct peaks are observed, corresponding to an off state (4 ns), neutral exciton state (51 ns), singly charged state (26 ns), and doubly charged state (18 ns). Note that the average-lifetime values are slightly shorter than the (un)charged exciton lifetimes determined in Fig. 4 of the main text, because the calculation of average lifetime includes some (un)charged biexciton photons.

**Supplementary Note 4.2:** Biexciton Cascades

The selection of cascade events from the photon stream can lead to false cascade detections on our SPAD array. First, despite a low dark count rate per pixel ($k_D < 100$ s$^{-1}$), the spectral dispersion over ~40 pixels increases the likelihood of cascade events containing at least one dark count. Second, pixels can re-emit photons after absorption, causing crosstalk detections on neighbouring pixels.[5,6] To reduce these sources of false positives, we apply time gating to the first and second photon of each cascade event based on the lifetime of the exciton and biexciton of each state. The first photon of a cascade event should be detected within approximately 3 times the (un)charged biexciton lifetime. For the BNC analysed in the main text, these are 33, 23 or 18 ns for the neutral, singly charged or doubly charged biexciton, respectively. Similarly, the second photon of the cascade should be detected within 3 times the (un)charged exciton lifetime, i.e. within 170, 84 or 60 ns after detection of the first photon. These upper limits of the time gates help us reject dark counts occurring over the full laser repetition period. We put a lower limit of 1 ns on the delay time between first and second photon to reject crosstalk events. After this time-gating procedure on the photon cascades, we can reconstruct the (un)charged-biexciton PL decay curves and emission spectra by analysing all first photons of the cascade events.

We scale the (un)charged-biexciton PL decay curves in the main text, such that the amplitude is proportional to the radiative decay rate. Three correction factors were applied:

1. Residence time (number of time bins) in the given charge state,

2. Emission efficiency for the second photon in the cascade (i.e., the (un)charged exciton emission efficiency), and

3. Loss of second-photon events due to the time-gating procedure on cascade events.

For (3), we estimated the fraction $f$ of second photons that were accepted by time gating with lower limit $t_1 = 1$ ns and and upper limit $t_2$ (170, 84, or 60 ns for the BNC in the main text):

$$f = \int_{t_1}^{t_2} \frac{1}{\tau} e^{-t/\tau} \, dt = e^{-t_1/\tau} - e^{-t_2/\tau}, \tag{S3}$$

where $\tau$ is the fitted lifetime of the (un)charged exciton state. Using these corrector factors, the (un)charged-biexciton PL decay curves show the expected trend of increasing radiative rate with increasing number of uncompensated charges in the BNC.

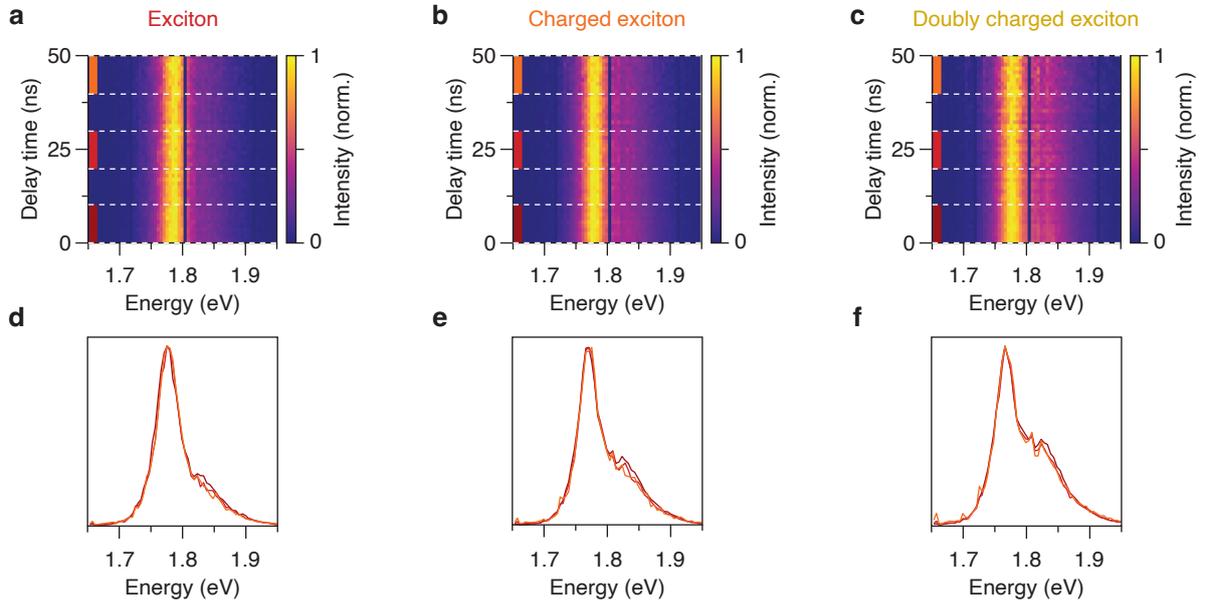

**Supplementary Fig. 6 | Time-resolved emission maps of the different charge states.** Normalised time-resolved emission maps showing the spectral evolution of **(a)** the neutral exciton, **(b)** the singly charged exciton, and **(c)** the doubly charged exciton, as a function of delay time. Spectra integrated over delay ranges of 0–10, 20–30 and 40–50 ns for **(d)** the neutral exciton, **(e)** the singly charged exciton, and **(f)** the doubly charged exciton. Over the first 10 ns, a slight decrease in the higher-energy P-level contribution is observed, which stabilises at longer decays. This behaviour is attributed to the presence of the (un)charged biexciton, which has a shorter lifetime than the (un)charged exciton and makes a contribution only at short times. At later delay times, when only exciton emission remains, the spectral shape remains unchanged. These results indicate that the decay from the two emitting levels is thermally coupled.

## Supplementary Note 4.3: Biexciton lifetimes

Because of the limited number of detected photons in the cascaded PL decay curves (~1000 counts), we again employ a maximum-likelihood estimation (MLE) fitting procedure to obtain the biexciton lifetimes $\tau_{BX}$. MLE provides a more accurate estimation than least-squares fitting in the low-count regime, as it accounts for the Poissonian nature of photon-counting statistics.[7,8]

We assume single-exponential decay:

$$I(t) = \frac{1}{\tau} \exp\left(-\frac{t}{\tau}\right) \tag{S4}$$

To account for the instrument response function (IRF), we convolve this decay with a Gaussian IRF with standard deviation $\sigma$ and offset $t_0$, yielding the analytical form:

$$I_{\text{conv}}(t; \tau) = \frac{1}{N} \exp\left(-\frac{t - t_0}{\tau}\right) \text{erfc}\left(\frac{\sigma^2 - (t - t_0)\tau}{\sqrt{2}\,\sigma\,\tau}\right) \tag{S5}$$

where erfc($t$) is the complementary error function, and $N$ is a normalisation factor:

$$N = \tau \left[ \exp\left(-\frac{\sigma^2}{2\tau^2}\right) + \mathrm{erfc}\left(\frac{\sigma}{\sqrt{2}\,\tau}\right) \right] \tag{S6}$$

Under Poisson statistics the negative log-likelihood function is given by:

$$\mathcal{L} = -\sum_i (y_i \ln I_i - I_i) \tag{S7}$$

where $I_i = I_{\mathrm{conv}}(t_i; \tau)$ is the expected number of photon counts at delay time $t_i$ according to the model and $y_i$ is the measured number of photons. We minimise $\mathcal{L}$ with respect to $\tau$ and scale the normalised model to match the total photon count in our histogram.

**Supplementary Note 4.4:** Exciton lifetimes

To disentangle the contributions of exciton and biexciton emission in the total time-averaged decay curves, we performed biexponential fits. The total decay is modeled as a sum of two single-exponential components: a fast decay attributed to biexciton recombination and a slower decay corresponding to exciton emission. We use the independently determined biexciton lifetimes for each charge state as fixed input parameters for the fast component in the biexponential fits. This constraint ensures that the extracted amplitude and lifetime of the slower component more accurately reflects the properties of the exciton. To minimise distortion from delayed emission due to reversible charge carrier trapping[8], we restricted the fit to the linear portion of the decay curve on a logarithmic scale. The fitting window was selected manually for each charge state through visual inspection. Specifically, we used the first 125 ns for the neutral state, 75 ns for the singly charged state, and 60 ns for the doubly charged state.

**Supplementary Note 4.5:** Emission efficiencies

In our experiments, the laser fluence is kept constant. As excitation occurs at high energy (405 nm), we assume that the absorption cross section is independent of charge state. The emission efficiency of a charged exciton state is therefore determined from its photon count rate (centre of a Gaussian fit to the count-rate histogram) relative to that of the neutral exciton, which has been determined to have unity efficiency with previous photonic experiments on single semiconductor particles.[4]

To evaluate the emission efficiency of the (un)charged biexciton cascade, we analyse the second-order correlation function $g^{(2)}$. We extract the biexciton-to-exciton emission ratio by fitting the correlation histogram with three double-sided exponentials with equal lifetimes, corresponding to the exciton lifetime, but allowing for independent amplitudes for the central (zero-delay) and side peaks. The data point at a delay time of to $\tau = 0$ is excluded from the fit, as our crosstalk correction introduces an artificial suppression of photon-pair detections at $\tau = 0$. We multiply the amplitude ratio from the fit with the emission efficiency of the (un)charged exciton state to obtain the emission efficiency of the (un)charged biexciton state.

**Supplementary Note 5:** Analysis of multi-particle spectroscopy

To extract spectral features from individual BNCs, we implemented a multistep correction and fitting procedure. First, we corrected for slight curvature in the spectrometer slit by centering each row of the EMCCD image. This alignment was based on a pixel shift derived from a reference measurement of the laser reflection at 405 nm spectrally dispersed on the camera. Following this, a flat background was subtracted from each row on the EMCCD image using the average signal in the leftmost 20 pixels, which lay outside the BNCs emission range.

To compare spectra recorded at different temperatures, we accounted for potential vertical shifts. This was achieved by shifting images with respect to each other to maximise the correlation between the summed intensity profiles of each image.

We then identified particle positions based on the high-temperature measurement, using both a minimum and maximum intensity threshold to select bright, well-isolated particles. This approach thus excluded BNCs that had photobleached or gone dark at elevated temperature.

To accurately describe the PL spectra of single BNCs, we fitted the data using a sum of three Gaussian functions with equal width ($\sigma$). While most BNCs exhibited two main peaks corresponding to the S and P exciton state, a subset displayed an additional, red-shifted emission peak. We attribute this third peak to emission from a trap state and excluded it from our analysis. For each selected BNC, we initially performed an unconstrained fit. From this set of preliminary fits, we computed the average $\sigma$ across all particles and used this as a fixed parameter in a second, constrained fit for each spectrum. This final fit yielded robust estimates of the energies of the two dominant transitions $E_\mathrm{S}$ and $E_\mathrm{P}$, their energy separation $\Delta E_\mathrm{P-S}$, and the intensity ratio $R = I_\mathrm{P}/I_\mathrm{S}$. To prevent bias due to the quenching of a subset of BNCs at higher temperatures, we consider only BNCs emissive at both temperatures for the analysis of Fig. 2 in the main text.

**Supplementary Note 6:** Single-particle quantum confinement model for exciton states

**Supplementary Note 6.1:** S–P confinement energy

In a spherical quantum dot, the confinement energy spacing can be approximated using the particle-in-a-box model. The S–P spacing scales inversely with the square of the confinement radius and inversely with the effective mass of the carrier. For electrons and holes confined over different volumes, the total S–P spacing in the emission spectrum is the sum of the electron and hole contributions:

$$\Delta E = \Delta E_e + \Delta E_h \propto \frac{1}{m_e R_e^2} + \frac{1}{m_h R_h^2} \tag{S8}$$

Here, $m_e = 0.1 m_0$ and $m_h = 1.0 m_0$ are the effective masses of the electron and hole, respectively. The confinement radii for a typical BNC from our batch are used $R_e = 7.7$ nm for the electron and $R_h = 5.05$ nm for the hole. By inserting these values, we obtain the separate contributions to the S–P level spacing for electrons and holes. These values are used directly in the modelling presented in later Supplementary Notes, where $\Delta E_e$ and $\Delta E_h$ enter the discrete and continuous Fermi–Dirac descriptions of multi-carrier state filling.

**Supplementary Note 6.2:** Degeneracy of bright exciton states

Including spin and valence-band multiplicity, there are 2×4 = 8 SS exciton configurations and 6×12 = 72 PP exciton configurations. However, only electron–hole pairs with matching orbital symmetry and orientation are optically bright (e.g., a $P_x$ electron recombining with a $P_x$ hole). This yields a bright PP fraction of 1/3, corresponding to an effective degeneracy ratio of

$$\frac{g_P}{g_S} = 3. \tag{S9}$$

We need this degeneracy ratio for our Boltzmann model (equation (1), main text), which describes the intensity ratio $R$ between bright SS and PP excitons as a function of $\Delta E_{P-S}$.

**Supplementary Note 6.3:** Relation between the energy-level spacing and the optical bandgap

To describe the experimentally observed correlation between $\Delta E_{P-S}$ and $E_S$ (Fig. 2d, main text) we use the particle-in-a-box model for quantum confinement. In this model, the energy of a confined carrier is given by:

$$E = \frac{\hbar^2 \chi_{n,l}^2}{2m^* R^2} \tag{S10}$$

where $\chi_{n,l}$ is the $n$-th zero of the spherical Bessel function of order $l$, $m^*$ is the effective mass of the carrier and $R$ is the radius of the confinement. We express $\Delta E_{P-S}$ in terms of $E_S$, using:

$$\Delta E_{P-S} = E_P - E_S = (\chi_{11}^2 - \chi_{10}^2) \frac{\hbar^2}{2} \left( \frac{1}{m_e R_e^2} + \frac{1}{m_h R_h^2} \right) \tag{S11}$$

and

$$E_S - E_g = \chi_{10}^2 \frac{\hbar^2}{2} \left( \frac{1}{m_e R_e^2} + \frac{1}{m_h R_h^2} \right) \tag{S12}$$

Taking the ratio of the two expression yields equation (2) of the main text, which is a simple relation between $\Delta E_{P-S}$ and $E_S$. While the absolute energies $E_S$ and $E_P$ depend on the asymmetric confinement of the electrons and holes in our BNCs, their difference $\Delta E_{P-S}$ scales linearly with $E_S - E_g$.

**Supplementary Note 6.4:** Size-dependent emission spectra

We model the expected PL spectra of the neutral exciton in particles with varying degrees of quantum confinement by considering thermal population of excited states using Boltzmann statistics, while obtaining the energy-level spacing based on the effective-mass model. The energy levels are calculated from the solutions to the Schrödinger equation in a spherical hard-wall potential. We use the first 40 zeros of the spherical Bessel functions $j_l$, with angular momentum quantum number $l \in [0,70]$ and radial quantum number $n \in [1,40]$. Each level's degeneracy is given by $2l+1$, reflecting the number of magnetic sublevels for a given $l$.

The energy of the lowest transition is calculated using an effective-mass model, assuming the electron is confined over the whole particle ($2R_e = 15.5$ nm), and the hole is confined to the core ($2R_h = 10.5$ nm). The electron and hole effective masses are set to $m_e = 0.1 m_0$ and $m_h = 1.0 m_0$, where $m_0$ is the free electron mass. This yields the energy for the lowest allowed optical transition in a particle with the average geometry from the synthesis batch:

$$E_S = E_g + \frac{\hbar^2 \pi^2}{2 m_e R_e^2} + \frac{\hbar^2 \pi^2}{2 m_h R_h^2} \tag{S13}$$

The energy difference between the SS and PP exciton for the particle of the average geometry is

$$\Delta E = \left( \frac{\hbar^2}{2 m_e R_e^2} + \frac{\hbar^2}{2 m_h R_h^2} \right) (\chi_{11}^2 - \pi^2) \tag{S14}$$

In Fig. 3 of the main text, we model particles with the average core and total size (see above), as well as particles with different sizes. Specifically, we consider particles with confinement energies scaled by factors 0.1, 0.6, 1, 2, 3, compared to the average particle. This represents particles ranging from bulk-like to strongly confined. For each case, we compute the Boltzmann-weighted populations of all 2840 levels and convolve them with Gaussian emission lines ($\sigma$ = 25.7 meV) centred at each transition energy to simulate the PL spectrum.

We compare the emission spectrum of the smallest simulated particle (topmost in Fig. 3d of the main text) with that obtained from a strongly confined model in which thermal excitation to higher energy levels is suppressed and thus all emission originates from the lowest-energy transition. The close agreement between the two spectra confirms that this particle is in the strong-quantum-confinement regime.

In contrast, the spectrum of the largest particle (bottommost in Fig. 3d of the main text) is compared with that predicted by a bulk model based on the joint density of states. Specifically, we solve for the electron chemical potential $\mu_e$ from the equation for Fermi–Dirac occupation, such that the total number of electrons in delocalization volume $V_e = \frac{4}{3}\pi R_e^3$ (with $R_e$ the total BNC radius) equals $n_e = 1$:

$$\frac{n_e}{V_e} = \int_0^\infty \frac{\sqrt{2m_e^3}}{\hbar^3 \pi^2} \sqrt{E}\, \frac{1}{e^{(E-\mu_e)/k_B T}}\, dE \tag{S15}$$

Similarly, the hole chemical potential $\mu_h$ is found for a delocalisation volume with core radius $R_h$ and $n_h = 1$ hole. Next, the predicted emission rate as a function of photon energy $E$ is

$$I(E) = A\sqrt{E} \times \frac{1}{1 + \exp\left[\left(\frac{m_h}{m_h + m_e}E - \mu_e\right)/k_B T\right]} \times \frac{1}{1 + \exp\left[\left(\frac{m_e}{m_h + m_e}E - \mu_h\right)/k_B T\right]} \tag{S16}$$

Here, $A$ is a prefactor that we scale to match the bulk and the quantum-confined thermal occupation models. The second factor on the righthand side describes the electron occupation at a vertical transition in the dispersion diagram at energy $E$ and the third factor the hole occupation at $E$. The predicted bulk emission spectrum, equation (S16), is also broadened by 25.7 meV.

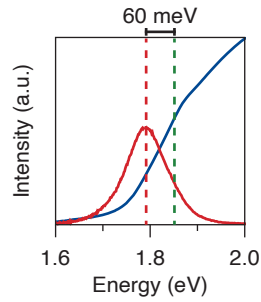

**Supplementary Fig. 7 | Quantifying the Stokes shift in bulk nanocrystals.** To qualitatively assess the deviation between our measured bandgap—obtained via temperature-dependent multiparticle spectroscopy—and literature values, we plot the ensemble photoluminescence (red) and absorption spectrum (blue). The Stokes shift is approximately 60 meV, consistent with the value of 1.69 eV for the bulk bandgap (Fig. 2d of the main text) extrapolated from the emission peak energies.

**Supplementary Note 7:** Fermi–Dirac occupation of electron and hole levels in multi-carrier states

The famous Fermi–Dirac distribution function

$$f(E) = \frac{1}{e^{(E-\mu)/k_BT} + 1} \tag{S17}$$

approximates fermion statistics in the limit of a large number of fermions, valid if the chemical potential $\mu$ changes negligibly upon addition of one more fermion.[9] The multi-carrier states studied in the main text, with at most 4 electrons and 2 holes, are not exactly in this limit. Nevertheless, we use this approximation in Fig. 4 of the main text and Supplementary Figs. 8a and b, where we need a distribution function for a continuous total number of electrons $n_e$ and holes $n_h$. For the modeling in Fig. 3 and Fig. 5, in contrast, we use a more exact description in terms of possible microstates of multi-carrier states with integer $n_e$ and $n_h$. The fermionic nature of electrons and holes is included by restricting the number of particles in S and P levels.

In this section, we will describe the discrete Fermi–Dirac occupation model and compare the predictions to those of Supplementary equation (S17) as well as to an even simpler Boltzmann model for electron and hole occupations.

**Supplementary Note 7.1:** Discrete microstate model

For a specific multi-carrier state with fixed total number of $n_e$ indistinguishable electrons and $n_h$ indistinguishable holes, we consider the distributions of electrons and holes over single-carrier levels separately.

The possible microstates for the electrons form a canonical ensemble with fixed $n_e$. The partition function is

$$Z = \sum_i g_i e^{-E_i/k_BT}. \tag{S18}$$

For our simple modeling in Fig. 3e of the main text, we consider only the S and P single-carrier levels. Each microstate $i$ is therefore defined by the number of S electrons $n_{e,S,i}$ and number of P electrons $n_{e,P,i} = n_e - n_{e,S,i}$. The energy of microstate $i$ is

$$E_i = n_{e,P,i} \Delta E_e, \tag{S19}$$

where the S-level energy is defined at zero and $\Delta E_e$ is the energy gap between conduction-band S and P levels (Supplementary Note 5). The degeneracy of microstate $i$ is

$$g_i = \binom{g_{e,S}}{n_{e,S,i}} \binom{g_{e,P}}{n_{e,P,i}}, \tag{S20}$$

where $g_{e,S} = 2$ and $g_{e,P} = 6$ are the degeneracies of the electron S and P single-carrier levels, respectively.

The fermion nature of electrons is included in the model by considering only microstates with $n_{e,S,i} \leq g_{e,S}$ and $n_{e,P,i} \leq g_{e,P}$. We obtain the expected number of S and P electrons, for a specific $n_e$, with

$$\langle n_{e,S} \rangle = \frac{1}{Z} \sum_i n_{e,S,i}\, g_i e^{-E_i/k_BT} \quad \text{and} \quad \langle n_{e,P} \rangle = n_e - \langle n_{e,S} \rangle. \tag{S21}$$

Equivalently, the expected numbers can be obtained by calculating the expected energy of the system and realizing that S electrons have zero energy by definition:

$$\langle n_{e,P} \rangle = \frac{k_B T^2}{\Delta E_e} \frac{\partial \ln Z}{\partial T} \quad \text{and} \quad \langle n_{e,S} \rangle = n_e - \langle n_{e,P} \rangle. \tag{S22}$$

The same procedure is used for the hole occupation of S and P levels for a total number of $n_h$, but using the valence-band degeneracies $g_{h,S} = 4$ and $g_{h,P} = 12$, representing the combined heavy- and light-hole bands.

## Supplementary Note 7.2: Comparison of the discrete-microstate model to the continuous Fermi-Dirac model and to a Boltzmann model

In Supplementary Fig. 8, we compare the outcome of the exact microstate model for fermion occupations (equation (S21) or, equivalently, equation (S22)) to the predictions of continuous Fermi–Dirac distribution (equation (S17)). The expected number of S and P electrons are calculated from the continuous Fermi–Dirac distribution by first solving $\mu$ from

$$g_S f(0) + g_P f(\Delta E) = n_e, \tag{S23}$$

using $f(E)$ from equation (S17), and then evaluating

$$\langle n_{e,S} \rangle = g_S f(0) \quad \text{and} \quad \langle n_{e,P} \rangle = g_P f(\Delta E). \tag{S24}$$

We see in Supplementary Fig. 8 that the two models produce very similar outcomes. The deviation is 10% (relative) for $n_e = 1$ and $n_e = 2$, and decreases as $n_e$ becomes larger. We accept this this level of deviation for the analysis in Fig. 4 of the main text and Supplementary Figs. 8a and b, because we need a continuous model to plot the rate constants as a function of $n_e$. In contrast, we use the more exact discrete model in Figs. 3,5 of the main text. For completeness, Supplementary Fig. 10 provides a side-by-side comparison of experiment, the continuous Fermi–Dirac model, and the discrete microstate model for the quantum efficiency, total lifetime, and P-state emission fraction.

Supplementary Fig. 8 also shows the outcome of a simple Boltzmann model, according to which

$$\langle n_{e,S} \rangle = \frac{g_S}{g_S + g_P e^{-\Delta E/k_B T}} \quad \text{and} \quad \langle n_{e,P} \rangle = n_e - \langle n_{e,S} \rangle. \tag{S25}$$

This model does not account for the fermion nature of electrons. Nevertheless, we see that the Boltzmann model reproduces the discrete model exactly for $n_e = 1$. This is intuitive, as the level occupations are in no way restricted by Pauli exclusion for $n_e = 1$. Indeed, only two microstates are possible for $n_e = 1$ (the electron occupies either an S or a P level) and the microstate degeneracies of equation (S20) reduce to the single-carrier level degeneracies. We thereby obtain equation (S25) exactly.

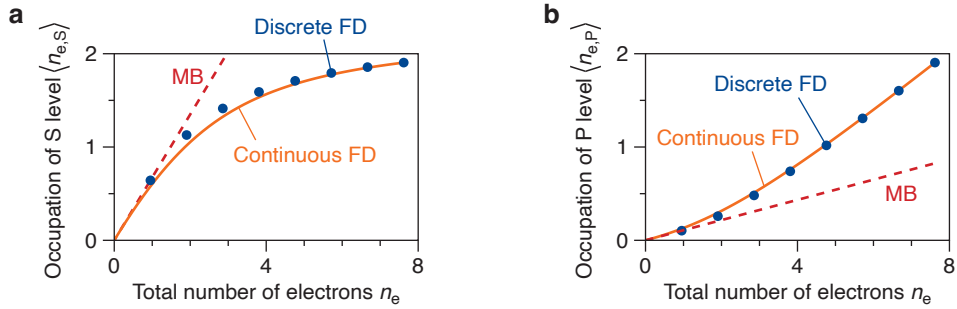

**Supplementary Fig. 8 | Level occupations according to different models. a,** The calculated occupation of the electron S level $\langle n_{e,S} \rangle$ as a function of the total number of electrons $n_e$ in a microstate, for an S–P energy gap of $\Delta E = 47$ meV. Blue datapoints: discrete Fermi–Dirac model (equation S21). Orange line: the well-known continuous Fermi–Dirac model (equations S17, S24). Red dashed line: the Boltzmann model (equation S25). **b,** Same as a, but for the occupation of the electron P level $\langle n_{e,P} \rangle$.

## Supplementary Note 7.3: Radiative recombination rates

Radiative recombination occurs between electron–hole pairs occupying levels of the same symmetry (S–S, $P_x$–$P_x$, etc.). The emission rate from each channel is therefore estimated as:

$$k_{\text{rad}}^{SS} = B' n_e^S n_h^S, \qquad k_{\text{rad}}^{PP} = \frac{1}{3} B' n_e^P n_h^P, \tag{S26}$$

with the total rate

$$k_{\text{rad}} = k_{\text{rad}}^{SS} + k_{\text{rad}}^{PP}. \tag{S27}$$

Here $B'$ is a universal scaling factor—which is equal for the SS and PP transitions if the oscillator strengths are equal—and the additional factor of 1/3 for the P-state recombination accounts for the requirement of matching orbital orientation (Supplementary Note 5.2).

Equation (S26) is applied to both models: (i) the continuous Fermi–Dirac model (Figs. 4e–g, main text), where $\langle n_{e,h}^{S,P} \rangle$ are continuous average occupations, and (ii) the discrete microstate model (Fig. 5, main text), where $\langle n_{e,h}^{S,P} \rangle$ are obtained from an average over microstates with fixed integer carrier numbers. In both cases, a single value of $B'$ is chosen such that the calculated total radiative decay rate of the uncharged biexciton matches the experimental value.

The hole degeneracies are relatively large, so the fermionic restrictions play only a minor role. As a result, the number of holes in the S and P levels for $n_h = 2$ is approximately twice that for $n_h = 1$ (to within 4% for S and 2% for P). We therefore divide the experimental $k_{\text{rad,S}}$ and $k_{\text{rad,P}}$ by the number of holes $n_h$ to collapse the data onto a single curve in Figs. 4f,g of the main text.

**Supplementary Note 7.4:** Auger recombination and quantum efficiency

To analyse non-radiative recombination, we assume that the dominant processes are Auger recombination through the positive and negative trion pathways. The prefactors $C^+$ and $C^-$ were obtained by fitting Eq. 9 of the main text to the experimental non-radiative decay rates. The values were then used to extrapolate Auger recombination rates to higher multi-carrier states up to $n_e = 5$ and $n_h = 5$.

The quantum efficiency is calculated as

$$Q = \frac{k_{\text{rad}}}{k_{\text{Auger}} + k_{\text{rad}}}, \tag{S28}$$

and the total recombination rate as

$$k_{\text{tot}} = k_{\text{Auger}} + k_{\text{rad}}, \tag{S29}$$

with the corresponding total lifetime

$$\tau = \frac{1}{k_{\text{tot}}}. \tag{S30}$$

The data provide no indications that the Auger recombination rates could depend on the distribution of electrons and holes over S and P levels. This is why our simple model for Auger recombination rates (Eq. 9 of the main text) depends only on the total number of electrons and holes. In Supplementary Fig. 8, we consider even simpler models, neglecting the difference between the positive- and negative-trion pathways and/or neglecting the discrete nature of the charge-carrier densities in the BNCs. The model of the main text (Eq. 9) is the best match to the experimental data.

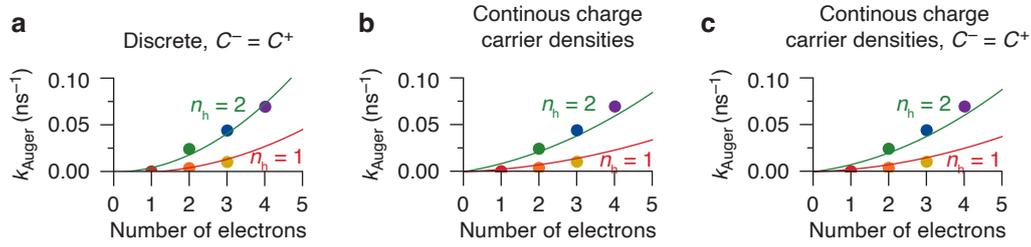

**Supplementary Fig. 9 | Alternative Auger models in finite-sized nanocrystals.** Total non-radiative decay rate as a function of the number of electrons $n_e$. Each panel shows the same experimental data as Fig. 4h of the main text, but the solid lines are fits to the data using different model assumptions. **a,** Finite, discrete charge carrier numbers are assumed, with equal rates for the negative-trion and the positive-trion Auger recombination pathways. The model follows $k_{nonrad} = C n_e n_h (n_e + n_h - 2)$, where the term –2 accounts for the finite number of charge carriers involved. **b,** Simpler model commonly used for bulk semiconductors that disregards the discrete charge carriers, but with asymmetry in the negative- and positive-trion Auger pathways: $k_{nonrad} = C^- n_e^2 n_h + C^+ n_e n_h^2$. **c,** Same as **b**, but now assuming equal rates for both Auger pathways: $k_{nonrad} = C n_e n_h (n_e + n_h)$. The model considered in the main text (Fig. 4h; considering finite number of charge carriers and asymmetric positive- and negative-trion Auger pathways) is a better match to the data than any of the models considered here. Specifically, the optimum least-squares loss function of the fit is worse than the for the main-text model by a factor 4 for the model of panel **a**, and by a factor 10 for the models of panels **b** and **c**.

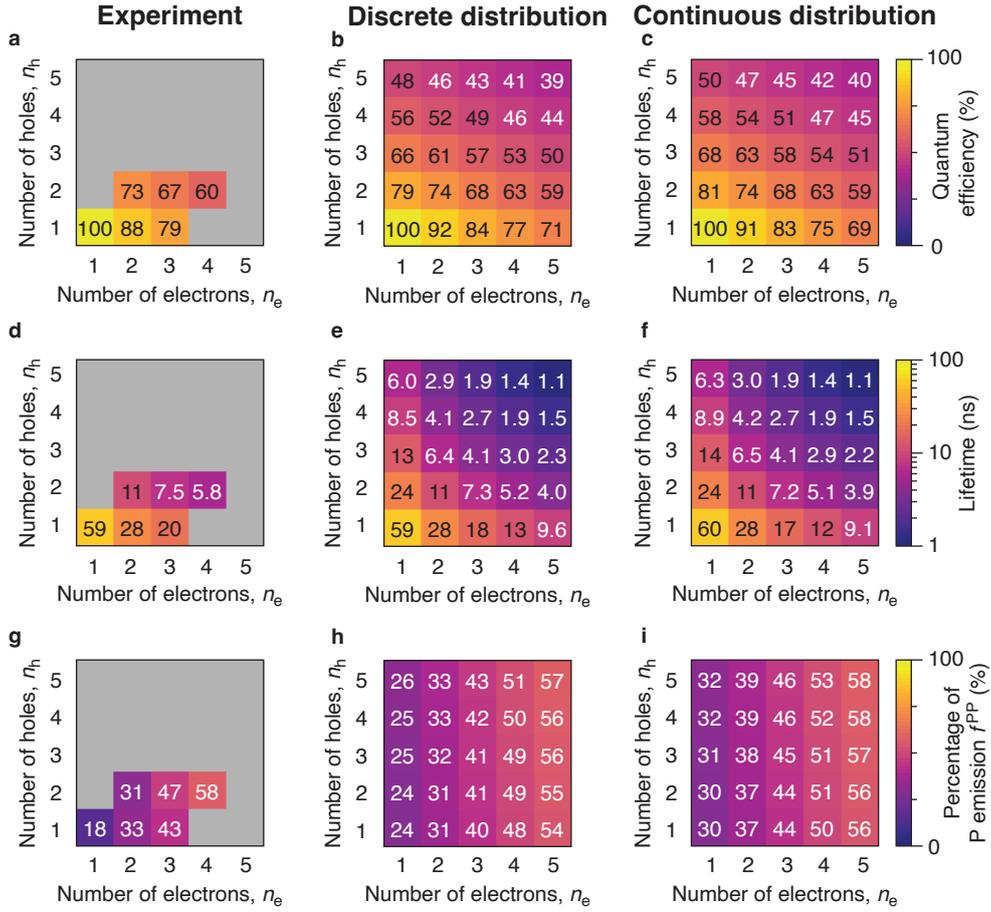

**Supplementary Fig. 10 | Comparison between predictions of the continuous and discrete distribution models. a,d,g,** Experimental values of **(a)** quantum efficiency, **(d)** excited-state lifetime, and **(g)** P-state emission fraction, as a function of $n_e$ and $n_h$ for the six multi-carrier states investigated. This is the same data as in Fig. 5 of the main text. **b,e,h,** Corresponding theoretical values using the discrete Fermi–Dirac distribution function (equation S21) and rate constants determined in Fig. 4 of the main text. These are the same calculation results as in Fig. 5 of the main text. **c,f,i,** Theoretical values using the continuous but approximate Fermi–Dirac distribution function (equation S17) together with the rate constants of Fig. 4 of the main text.

# Extended Data

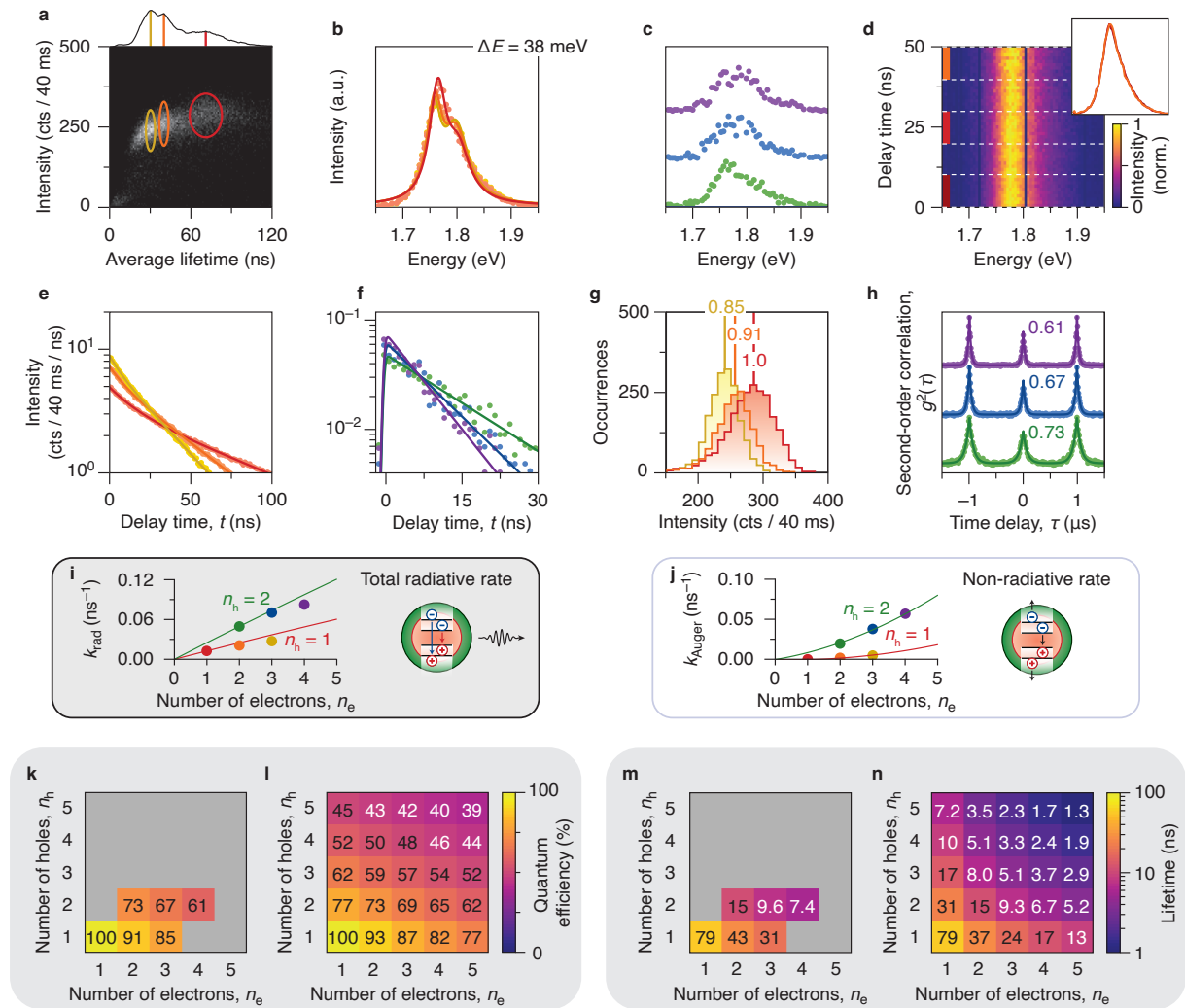

**Supplementary Fig. 11 | Summary of the experimental results for a second BNC of the same synthesis batch. a,** Fluorescence-lifetime–intensity distribution (FLID), showing fluctuations in average lifetime and intensity across 40-ms time bins, caused by spontaneous charging events in the BNC. Three distinct emissive states are identified. These are indicated in the FLID and in the lifetime histogram (top): neutral (red), singly negatively charged (orange), and doubly negatively charged. **b,** Reconstructed PL spectra of the neutral exciton (red), singly charged exciton (= trion; orange), and doubly charged exciton (yellow). **c,** Same as **b**, but for the neutral biexciton (green), singly charged biexciton (blue), and doubly charged biexciton (purple). We observe a rise in the relative intensity of the higher-energy peak with the number of excess electrons, similar to the BNC of the main text. This reflects increased population of the electron P level. **d,** Normalised time-resolved emission map showing spectral evolution of the neutral exciton emission as a function of decay time. The inset displays the average spectra for delay ranges of 0–10 ns, 20–30 ns, and 40–50 ns. The constant shape of the spectrum over time shows thermal equilibrium between the two emissive states. **e,** PL decay curves of the neutral (red), singly charged (orange), and doubly charged (yellow) states. Exciton lifetimes are extracted from the slow component of a biexponential fit. **f,** PL decay curves of the neutral (green), singly charged (blue), and doubly charged (purple) biexciton states, reconstructed by analysing photon cascades. **g,** Histogram of photon count rates from the neutral (red), singly charged (orange), and doubly charged (yellow) states. The solid lines show the centre of a Gaussian fit. Assuming a quantum efficiency of 1.0 for the neutral state, we estimate the efficiencies of the other states from the relative count rates.[4] **h,** Second-order correlation functions of the neutral (green), singly charged (blue), and doubly charged (purple) state of the BNC. We estimate the biexciton quantum efficiencies from the peak height in the correlation function and the exciton quantum efficiencies determined in panel **g**. **i,** The total radiative decay rate as a function of number of electrons $n_e$. The total radiative rate follows the expected statistical scaling (equation (5) of the main text) for one and two electrons but deviates slightly from the model at higher $n_e$. We attribute this to a reduced electron–hole wavefunction overlap due to increased electron–electron repulsion at higher occupation. **j,** The total non-radiative decay rate as a function of $n_e$. The solid lines represent the expected non-radiative rates for the various charge configurations (equation (9) of the main text) for Auger coefficients $C^- = 0.0009$ and $C^+ = 0.0038$ ns$^{-1}$ for the negative- and positive-trion Auger pathways, respectively. **k,** Experimental quantum efficiencies of multi-carrier states as a function of the number of electrons $n_e$ and the number of holes $n_h$. **l,** Same, but calculated using the model of equations (8) and (9) (of the main text) and using the parameters determined in **i, j**. The model is extrapolated to multi-carrier states with up to 5 electrons and 5 holes. **m,n,** Same as **k,l**, but showing the lifetime of multi-carrier states.